\documentclass[prd,preprintnumbers,twocolumn,eqsecnum,floatfix,a4paper,nofootinbib,superscriptaddress]{revtex4}
\usepackage{color}
\usepackage{calc}
\usepackage{amsmath,amssymb,graphicx}
\usepackage{amssymb,amsmath}
\usepackage{bm}
\usepackage{microtype}
\usepackage{booktabs}
\usepackage{times}
\usepackage{subfigure}
\usepackage[varg]{txfonts}
\usepackage[utf8]{inputenc}
\definecolor{LinkColor}{rgb}{0.75, 0, 0}
\definecolor{CiteColor}{rgb}{0, 0.5, 0.5}
\definecolor{UrlColor}{rgb}{0, 0, 0.75}
\allowdisplaybreaks
\DeclareFontFamily{OT1}{pzc}{}
\DeclareFontShape{OT1}{pzc}{m}{it}{<-> s * [1.10] pzcmi7t}{}
\DeclareMathAlphabet{\mathpzc}{OT1}{pzc}{m}{it}
\usepackage[breaklinks, plainpages=false, colorlinks=true, anchorcolor=cyan, linkcolor=red, citecolor=blue, urlcolor=magenta, bookmarks=false]{hyperref}
\usepackage{natbib}
\usepackage{blindtext}
\begin{document}
\renewcommand{\thefigure}{\arabic{figure}}
\setcounter{figure}{0}
\def\I{{\rm i}}
\def\E{{\rm e}}
\def\D{{\rm d}}

\title{Acceleration Relations in the Milky Way as Differentiators of Modified Gravity Theories}

\author{Tousif Islam}
\email[]{tousifislam24@gmail.com}
\affiliation{ Center for Scientific Computation and Visualization Research (CSCVR), University of Massachusetts (UMass) Dartmouth, Dartmouth, MA-02740, USA}
\affiliation{ International Centre for Theoretical Sciences, Tata Institute of Fundamental Research, Bangalore- 560012, India}

\author{Koushik Dutta}
\email[]{koushik.physics@gmail.com}
\affiliation{Department of Physical Sciences, Indian Institute of Science Education and Research Kolkata, Mohanpur - 741 246, WB, India\\ Theory Divison, Saha Institute of Nuclear Physics \footnote{on lien to IISER, Kolkata}, HBNI,1/AF Bidhannagar, Kolkata- 700064, India}
\begin{abstract}
The dynamical mass of galaxies and the Newtonian acceleration generated from the baryons have been found to be strongly correlated. This correlation is known as `Mass-Discrepancy Acceleration Relation' (MDAR). Further investigations have revealed a tighter relation - `Radial Acceleration Relation' (RAR) - between the observed total acceleration and the (Newtonian) acceleration produced by the baryons. So far modified gravity theories have remained more successful than $\Lambda$CDM to explain these relations. However, a recent investigation has pointed out that, when RAR is expressed as a difference between the observed acceleration and the expected Newtonian acceleration due to baryons (which has been called the `Halo acceleration relation or HAR'), it provides a stronger test for modified gravity theories and dark matter hypothesis. Extending our previous work  \citep{kt2018}, we present a case study of modified gravity theories, in particular Weyl conformal gravity and Modified Newtonian Dynamics (MOND), using recent inferred acceleration data for the Milky Way. We investigate how well these theories of gravity and the RAR scaling law can explain the current observation.
\end{abstract}
\maketitle
\section{\textbf{Introduction}} 
In Newtonian gravity, i.e. the weak-field limit of the general relativity, the discrepancy between the mass estimated from the observed dynamics of galaxies ($M_{dyn}$) and the observed baryonic mass ($M_{bar}$) has been found to be correlated with the observed acceleration ($a_{obs}$) in the galaxy, showing a monotonous decline with increasing radial distances (or decreasing observed acceleration). The observed relation between $M_{dyn}/M_{bar}$ and $a_{obs}$ is known as Mass-Discrepancy-Acceleration Relation (MDAR) \citep{mcgaugh2004mass}. \\

Analyzing the high precision data from 153 spiral galaxies in SPARC (Spitzer Photometry and Accurate Rotation Curves) database, McGaugh, Lelli  and Schombert (MLS) \citep{McGaugh} have found a even tighter correlation between the radial acceleration, $a_{obs}$, inferred from the rotation curves and that expected Newtonian (centripetal) acceleration generated by the baryons in galaxies. The emperical relation, known as Radial Acceleration Relation (RAR), is quite similar to the acceleration law of Modified Newtonian Dynamics (MOND) \citep{mond1,famaey2012modified} and is given by:
\begin{equation}
a_{MLS}=\frac{a_{new}^{bar}}{1-exp(-(\frac{a_{new}^{bar}}{a{\dagger}})^{1/2})},  
\label{master}
\end{equation}
where $a_{new}^{bar}$ is the Newtonian acceleration produced by the baryonic mass only and $a{\dagger}=1.2 \times 10^{-10}$ $m s^{-2}$ is the acceleration scale. Lelli \textit{et al.} \cite{lelli2017one} have further established that similar relation holds for other types of galaxies such as ellipticals, lenticulars, and dwarf spheroidals. The universality of RAR across different types of galaxies along with its small scatter provides an unique test for dark matter models and modified gravity theories at galactic scale. Even though semi-analytical dark matter models can account for the RAR, the intrinsic scatter produced by these models is always significantly larger than the one observed \citep{di2015mass,desmond2016statistical}. Furthermore, within the context of $\Lambda$CDM where dark matter dominates the baryonic mass, it is not immediately clear why the observed acceleration should be strongly correlated to the baryonic matter. It is thus natural to investigate whether the existence of such scaling could be a hint for modification of gravity at the galactic scales. Modified gravity theories such as Modified Newtonian Dynamics (MOND) \citep{mond1,famaey2012modified}, Weyl Conformal gravity \citep{weyl1,weylrot5} and Scalar-Tensor-Vector Gravity (STVG)/Modified Gravity (MOG) \citep{mog} have been shown to be in excellent agreement with RAR (\cite{ghari2019radial} for MOND; \cite{o2019radial,kt2018} for Weyl gravity; \cite{green2019modified} for MOG). However, \cite{lelli2017testing} found Emergent gravity \citep{verlinde2017emergent} to be inconsistent with RAR.\\

Tian and Ko \cite{tian2019halo}, on the other hand, found that expressing RAR in terms of the difference between the observed acceleration and the expected Newtonian acceleration due to baryons (which they call as `halo acceleration') provides more interesting features:
\begin{equation}
a_{h}=a_{obs}-a_{new}^{bar}.
\end{equation}
They claim that the halo acceleration ($a_{h}$), when plotted as function of the expected Newtonian acceleration due to baryons, shows a prominent maxima. They further observed that HAR provides a much stringent test for different astrophysical dark matter profiles and different versions of MOND (with different interpolating functions). \\

We note that RAR have been obtained by fitting the cumulative (inferred) acceleration data of hundreds of galaxies \citep{McGaugh}. However, the obtained relation has also been tested individually for the galaxies in the SPARC catalog \citep{li2018fittingrar}. The reported relation has been found in all types of galaxies irrespective of whether the corresponding data fall in the low acceleration regime  ($10^{-10}$ $m/s^{2}$ - $10^{-12}$ $m/s^{2}$) or in the high end ($10^{-8}$ $m/s^{2}$ - $10^{-10}$ $m/s^{2}$). HAR, on the other end, have not been fitted to individual galaxies so far. In this paper, we present an interesting case study of RAR and HAR in the Milky Way through the lens of modified gravity theories, namely Weyl conformal gravity and MOND. The Milky Way is one of the very few individual galaxies for which the rotation curve data allows one to probe both the high and low acceleration domain (from $10^{-8}$ $m/s^{2}$ to $10^{-12}$ $m/s^{2}$). Several groups [Sofue (YS12) \citep{sofue2};  Bhattacharjee et al (BCK14) \citep{pijush}; Huang et al (YH16) \citep{yh16}] have constructed highly resolved rotation curve for the Milky Way extending up-to a large galactocentric distance beyond $\sim$ 100 kpc using kinematical data of different types tracer objects, without assuming any particular model for the galaxy mass profile. \\

In our previous work \citep{kt2018} (DI18), we have complied the rotation curve data of YS12, BCK14 and YH16 and showed that both Weyl conformal gravity and MOND can reasonably fit the data. Extending the analysis done in KT18, we now use the inferred centripetal acceleration data to address the following questions: (1) Do the rotation curve data of the Milky Way follow MDAR, RAR and HAR? (2)  If yes, how well Weyl conformal gravity and MOND can explain these two phenomenological relations in the Milky Way? (3) Which of these three relations gives a stronger test for modified gravity theories ? Our paper is organized in the following way. We first present the mass model of the Milky Way in Section \ref{sec2}; then provide a brief description of the Weyl Conformal gravity and MOND in Section \ref{sec3}; discuss our results in Section \ref{sec4}; and finally pen down the summary in Section \ref{sec5}.
\section{\textbf{Milky Way Mass Profile}} 
\label{sec2}
Following \cite{mcmillan}, we model the Milky Way (MW) galaxy with five distinct structural components: a spherical central bulge, thin and thick stellar disks, and HI and molecular gas disks. The central bulge is assumed to follow an exponential surface brightness profile \citep{bulge} which is translated into the following three dimensional mass density
\begin{equation}
\rho(r)=\frac{M_{bulge}}{2\pi^2 t^3}K_0(r/t),
\label{eq1}
\end{equation}
where $M_{bulge} = 2.0 \pm 0.3 \times 10^{10} M_{\odot}$  is the total mass of the bulge \citep{bulge2}, $t$ is the extent of the bulge and $K_0$ denotes modified Bessel function. The exact value of $t$ remains uncertain in literature (ranging from 0.6 kpc to 2.0 kpc). Here, we use an average value of $t=1$   kpc. For the disk components, we use usual exponential surface mass density profiles of the form
\begin{equation}
\Sigma(r)=\Sigma^{0} e^{-r/R},
\label{eq2}
\end{equation}
where  $\Sigma$, $\Sigma^{0}$ and $R$ are the surface mass density, maximum surface density (at the center) and the scale length of the disk respectively. For different disk components (thin stellar disk/ thick stellar disk/ HI disk / H2 molecular gas disk), $\Sigma$, $\Sigma^{0}$ and $R$ would take different values (Table \ref{T1}). Apart from these, we include a central super-massive black hole with a mass $M_{bh} = 4.0\pm0.3\times10^{6} M_{\odot}$ in the mass model.
\begin{figure*}
	\begin{center}
		\includegraphics[scale=0.55]{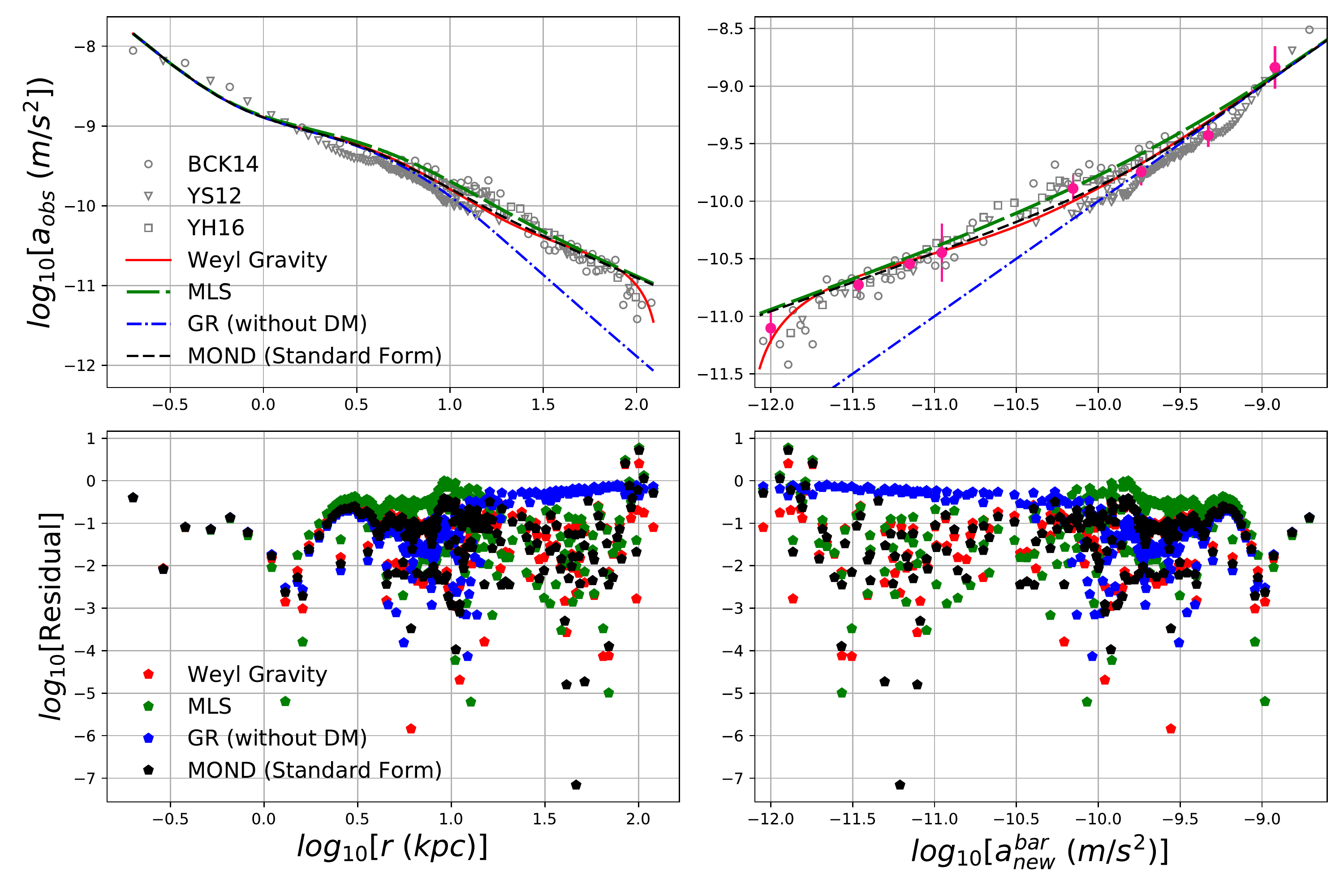} 
	\end{center} 
	\caption{\textbf{\textit{Upper Left: } Observed centripetal acceleration (inferred from YH12, YS17  and  BCK14)  as  a  function  of  radial  distances from the galactic center in log-log scale. \textit{Upper Right: } loglog plot of observed centripetal acceleration as a function of Newtonian expectation due to baryons. Predicted profiles in GR (without dark matter), Weyl gravity, MOND and RAR scaling given in Eq. \ref{master} by McGaugh, Lelli  and Schombert (MLS) \citep{McGaugh} are then superimposed in both panels. Binned data has been plotted in pink circles. \textit{Lower Left: } Residuals [for GR (without dark matter), Weyl gravity, MOND and RAR scaling] as a function of radial  distances from the galactic center in log-log scale. \textit{Lower Right: } Residuals as a function of Newtonian expectation due to baryons. Color codes are given in the legend. Details are in the texts.}}
	\label{fig1}
\end{figure*}
\begin{table}[b]
	\centering
	\caption[Milky Way mass model]{\textbf{Parameters for the Milky Way mass model \cite{mcmillan}}}
	\label{T1}
	\begin{tabular}{c c c}
		\hline 
		\hline
		&$\Sigma_0$ &$R$\\
		Thin Stellar Disk &$ 886.7 \pm 116.2$ $ M_{\odot} pc^{-2}$ &2.6 $\pm$ 0.52 kpc\\
		Thick Stellar Disk &$ 156.7 \pm 58.9$ $ M_{\odot} pc^{-2}$ & 3.6  $\pm$ 0.72 kpc\\
		HI Disk & $ 1.1 \times 10^{10} M_{\odot}$  & 7.0 kpc\\
		H2 Disk & $ 1.2 \times 10^{9} M_{\odot}$ & 1.5 kpc\\
		\hline 
		\hline
	\end{tabular} 
\end{table}
\section{Modified Gravity Theories}
\label{sec3}
\subsection{Weyl Conformal Gravity} Weyl conformal gravity \citep{weyl1,weylrot5} employs the principle of local conformal invariance of the space-time in which the action remains invariant under conformal transformation i.e. $g_{\mu \nu} (x) \rightarrow \Omega^{2}(x) g_{\mu \nu} (x)$, where  $g_{\mu \nu}$ is the metric tensor and $\Omega(x)$ is a smooth positive function.  It also obeys the general coordinate invariance and the equivalence principle. These requirements lead to a unique action $I_{w}=-\alpha_{g} \int d^{4}x \sqrt{-g} C_{\lambda\mu\nu\kappa} C^{\lambda\mu\nu\kappa}$ where $\alpha_{g}$ is a dimensionless coupling constant and $C_{\lambda\mu\nu\kappa}$  is the Weyl tensor \citep{weyl1918}. The action then yields a fourth order field equation. Mannheim and Kazanas have reported an exact vacuum solution for static, spherically symmetric geometry \citep{weylrot5}. \\
\indent It has been shown that, in Weyl gravity, the potential within a galaxy is decided by both the local mass distribution in the galaxy as well as the mass exterior to it \citep{weylrot5}. The global contribution to the potential has two different origins: the homogeneous cosmological background, contributing a linear potential, and the inhomogeneities in the form of galaxies, clusters and filaments, contributing a negative quadratic potential. \\ 
\indent In Weyl gravity, each star generates a potential $V^{\ast}_{star}(r > r_{0}) = - \frac{\beta^{\ast}c^{2}}{r} + \frac{\gamma^{\ast}c^{2}r}{2}$. Therefore, the potential in a disk component would be the summation of potentials generated by all such stars in the disk. The total contribution to rotational velocities of stars from the luminous mass within the disk  following a exponential surface mass density profile (Eq. (\ref{eq2}) is then found to be \citep{weylrot5}
\begin{align}
& 
\begin{aligned}[t]
&v^{2}_{disk}(r)\\
&=\frac{N\beta^{\ast}c^{2}r^{2}}{2R^{3}_{0}} \Big[ I_{0}\left(\frac{r}{2R_{0}}\right)K_{0}\left(\frac{r}{2R_{0}}\right) -I_{1}\left(\frac{r}{2R_{0}}\right)K_{1}\left(\frac{r}{2R_{0}}\right) \Big]\\
& + \frac{N\gamma^{\ast}c^{2}r^{2}}{2R_{0}} I_{1}\left(\frac{r}{2R_{0}}\right)K_{1}\left(\frac{r}{2R_{0}}\right),
\end{aligned}
\label{m4}
\end{align}
where $I_0$, $I_1$, $K_0$ and $K_1$ are modified Bessel functions and $N=2\pi\Sigma_0 R^{2}_{0}$ is the total number of stars \citep{weylrot5}. We note that the first term in Eq.  (\ref{m4}) is the contribution from the Newtonian term (or in GR; weak gravity limit), the second term originates from the linear potential. On the other hand, spherical bulge with mass profile similar to the one in Eq.  (\ref{eq1}) yield circular velocities of the form \citep{weylrot5}
\begin{align}
& 
\begin{aligned}[t]
&v^2_{bulge}(r)\\
&={2 N\beta^* c^2 \over \pi r} \int_0^{r/t}dz\, z^2K_0(z)+{N\gamma^* c^2 r\over \pi} \int_0^{r/t}dz\, z^2K_0(z)\\
&- { N\gamma^* c^2 t^2 \over 3 \pi r} \int_0^{r/t}dz\, z^4K_0(z)+{2 N\gamma^* c^2 r^3 \over 3\pi t^2}K_1(r/t).
\end{aligned}
\label{m7}
\end{align}
The first term denotes the contribution from the Newtonian potential whereas the second term is the Weyl gravity correction from the linear term. The rotational velocity for the Milky Way galaxy due to the local mass distribution is thus obtained as 
\begin{align}
v^2_{loc}(r)=&
\begin{aligned}[t]
&v^2_{bulge}(r) + v^2_{disk,thin}(r) + v^2_{disk,thick}(r)\\
&+ v^2_{disk,HI}(r) + v^2_{disk,H2}(r).
\end{aligned}
\label{m7}
\end{align}
Finally, we include the global effects and write down the net rotational velocity in Weyl  gravity \citep{weylrot5}:
\begin{eqnarray}
v^2_{tot}(r) = v^2_{loc}(r) + \frac{\gamma_0 c^{2} r}{2} - \kappa c^{2} r^{2}.
\label{m9}
\end{eqnarray}
The corresponding centripetal acceleration is thus : $\frac{v^2_{tot}(r)}{r}$. The values of the four universal Weyl gravity parameters are fixed by previous fits to the rotation curves of $\sim$ 100 galaxies \citep{weylrot1,weylrot2,weylrot3}: $\beta^{\ast} = 1.48 \times 10^{5}$ $cm$; $\gamma^{\ast} = 5.42 \times 10^{-41}$ $cm^{-1}$; $\gamma_0 = 3.06 \times 10^{-30}$ $cm^{-1}$ and $\kappa = 9.54 \times 10^{-54} $ $cm^{-2}$. These values have also been used in our previous study \cite{kt2018} of Weyl conformal gravity at galactic and extra-galactic scales. To maintain consistency, same choices have been made for the parameter values in this work.\\

It is, however, important to point out that, in Weyl gravity, each star generates a potential that consists of a Newtonian term plus a linearly growing term. We fix the coefficients of the Newtonian and linear terms to the values obtained from previous study of $\sim$100 galaxy rotation curves \citep{weylrot1,weylrot2,weylrot3} which considered the coefficients as free parameters that can be fitted to improve the model's agreement with data. However, it can be shown that if the matter source is a simple 3-dimensional delta function, the coefficient of the Newtonian term is zero and the entire potential is a linear term. The Newtonian term only acquires a non-zero coefficient if the matter source has a second derivative of a delta function. This yields results which are wildly inconsistent with the data. Attempts should be made to explore this direction further and find out ways to reconcile with data.
\subsection{Modified Newtonian Dynamcies (MOND)}In Modified Newtonian Dynamcies (MOND) \citep{mond1,famaey2012modified} scenarios, net acceleration is obtained via modifying the Newtonian acceleration due to baryons through an interpolating function $\mu$ such that
\begin{equation}
\mu \left(\frac{a}{a_0}\right) a = a_{N},
\end{equation}
$a_0$ denotes a critical value below which Newtonian gravity breaks down. The interpolating function $\mu(x) \approx x$ when $x \ll 1$ and $\mu(x) \approx 1$ when $x \gg 1$. Therefore, in MOND, Newtonian behavior is recovered when the acceleration is high. In literature,  different functional forms of the interpolating function $\mu(x=\frac{a}{a_0})$ is used. In this paper, we stick to the `standard' form: 
\begin{equation}
\mu (x) = \frac{x}{\sqrt{(1 + x^{2})}},
\end{equation}
with $a_0$ = $1.21 \times 10^{-10} m/s^{2}$. Therefore, the MOND acceleration can be written as \citep{mond1}
\begin{equation}
a_{MOND} =  \frac{a_{N}}{\sqrt{2}}\Big[ 1 +  \Big( 1 + \left(\frac{2a_0}{a_{new}^{bar}}\right)^2 \Big)^{1/2} \Big]^{1/2},
\end{equation}
where $a_{new}^{bar}$ is the Newtonian acceleration associated with the baryonic mass.
\begin{figure}
	\begin{center}
		\includegraphics[scale=0.6]{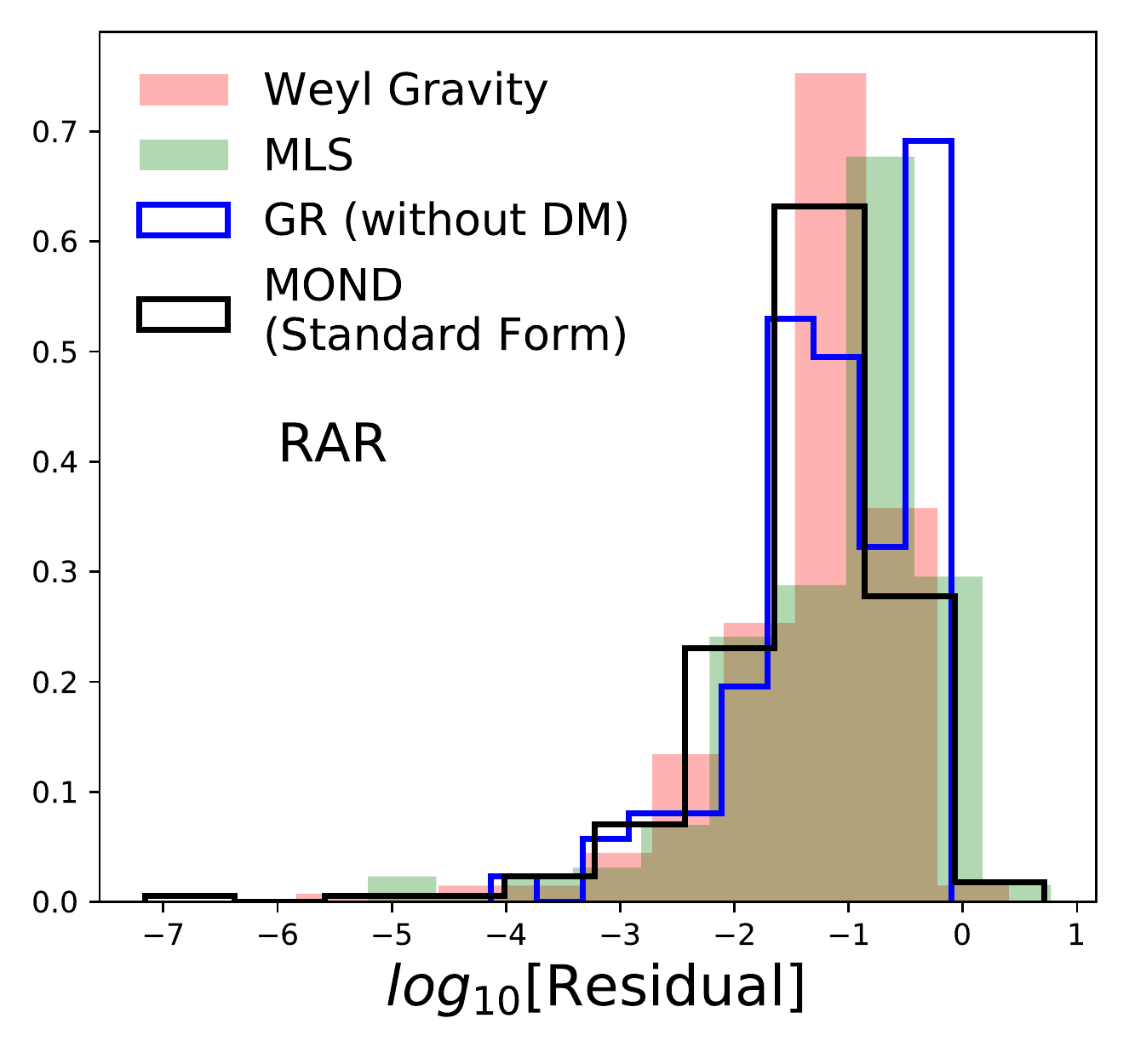} 
	\end{center} 
	\caption{\textbf{Histogram of the residuals between the inferred centripetal accelerations and the predicted accelerations in GR (without DM), Weyl gravity, MOND and RAR scaling. Color codes are given in the legend. Details are in the texts.}}
	\label{fig2}
\end{figure}
\section{\textbf{Results}}
\label{sec4}
\subsection{Radial Acceleration Relation (RAR) and Modified gravity}
\label{sec4a}
We first plot the inferred acceleration data for the Milky Way [obtained from BCK14 (49 data points), YS12 (123 data points) and YH16 (43 data points)] as a function of radial distances from the galactic center in Figure \ref{fig1} (upper left). As mentioned before, the acceleration data covers both the low acceleration regime  ($10^{-10}$ $m/s^{2}$ - $10^{-12}$ $m/s^{2}$) and high acceleration regime ($10^{-8}$ $m/s^{2}$ - $10^{-10}$ $m/s^{2}$). In particular, we find no noticeable feature in the transition zone from high to low acceleration regime. On top of the data, we superimpose the acceleration profile predicted in GR (blue dashed dotted), Weyl gravity (solid red line) and MOND (black dashed line). Furthermore, we show the expected profile when RAR scaling law (\cite{McGaugh}; referred to as MLS) is assumed to be valid (long dashed green line). No dark matter is assumed. We find that Weyl gravity, MOND and RAR (otherwise mentioned as MLS in the figure) overall match  with the data. However, the GR (without dark matter) profile departs from the data beyond $\sim$ 10 kpc from the galactic center. Interestingly, at $\sim$ 10 kpc, the acceleration reaches the value $\sim$  $10^{-10}$ $m/s^{2}$ which corresponds to the acceleration scale $a_0$ in MOND.\\

In Fig \ref{fig1} (upper right), we also plot the observed centripetal acceleration as a function of the expected Newtonian acceleration from baryonic matter only. On top of that, we plot the binned data for radial acceleration in pink circles. We note the following points. First, phenomologically established RAR can reasonably account for the observed data. This is not a surprise as the relation have been tested for a number of galaxies and is found to be quite robust. Though the overall shape of the MOND and Weyl gravity profiles differ a bit, both agrees to the data with comparable chi-square value (Table \ref{T2}). However, one can see that MOND overshoots the data in the extreme low end of the acceleration while both MOND and Weyl gravity shows slight disagreement in the extreme high end of the acceleration.\\
\begin{figure*}
	\begin{center}
		\includegraphics[scale=0.55]{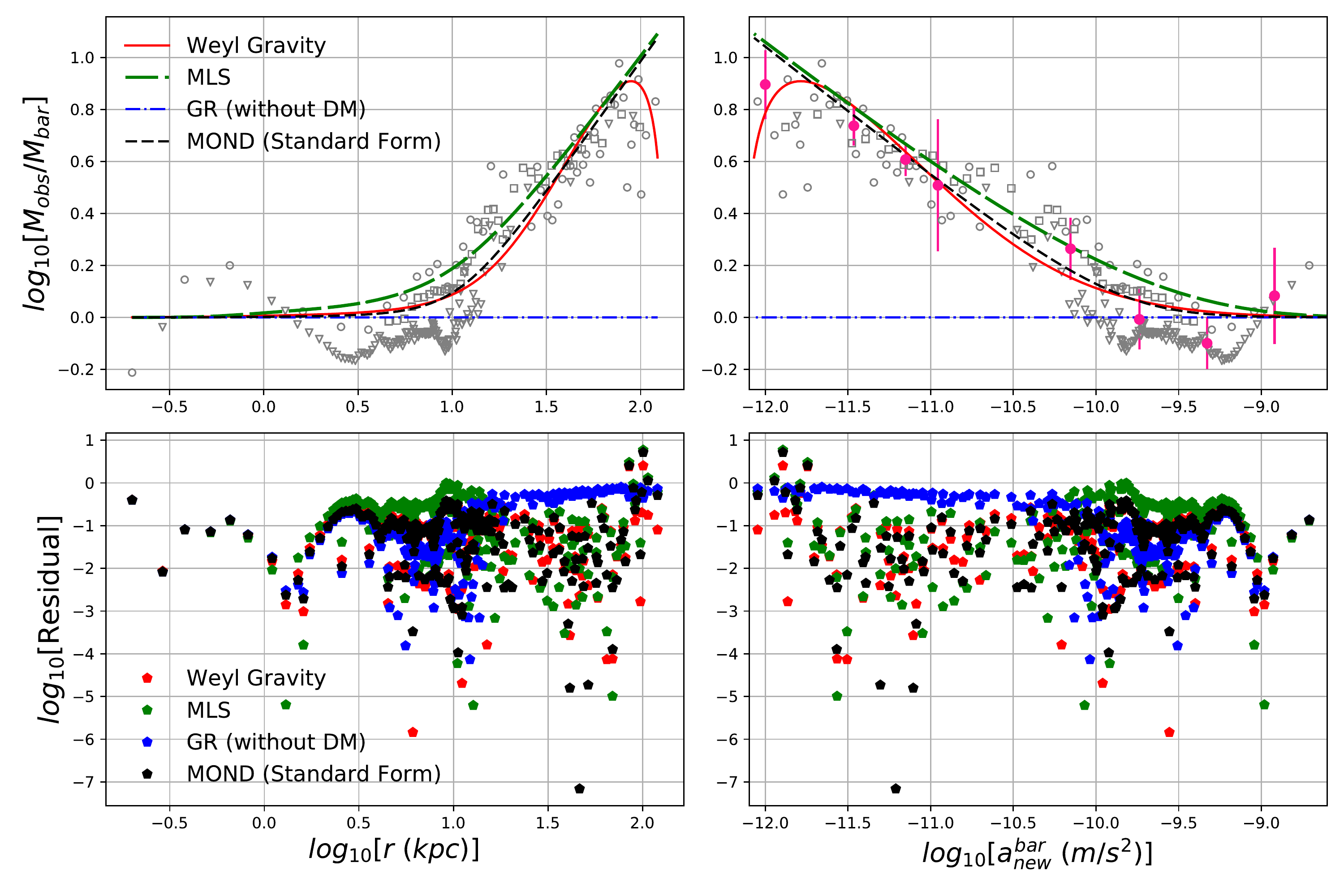} 
	\end{center} 
	\caption{\textbf{\textit{Upper Left: } Inferred mass discrepancy as  a  function  of  radial  distances from the galactic center in log-log scale. \textit{Upper Right: } loglog plot of inferred mass discrepancy as a function of Newtonian expectation due to baryons. Binned data has been plotted in pink circles. Predicted profiles in GR (without dark matter), Weyl gravity, MOND and RAR scaling are then superimposed in both panels. \textit{Lower Left: } Residuals [for GR (without dark matter), Weyl gravity, MOND and RAR scaling] as a function of radial  distances from the galactic center in log-log scale. \textit{Lower Right: } Residuals as a function of Newtonian expectation due to baryons. Color codes are given in the legend. Details are in the texts.}}
	\label{fig3}
\end{figure*}
\begin{figure}
	\begin{center}
		\includegraphics[scale=0.6]{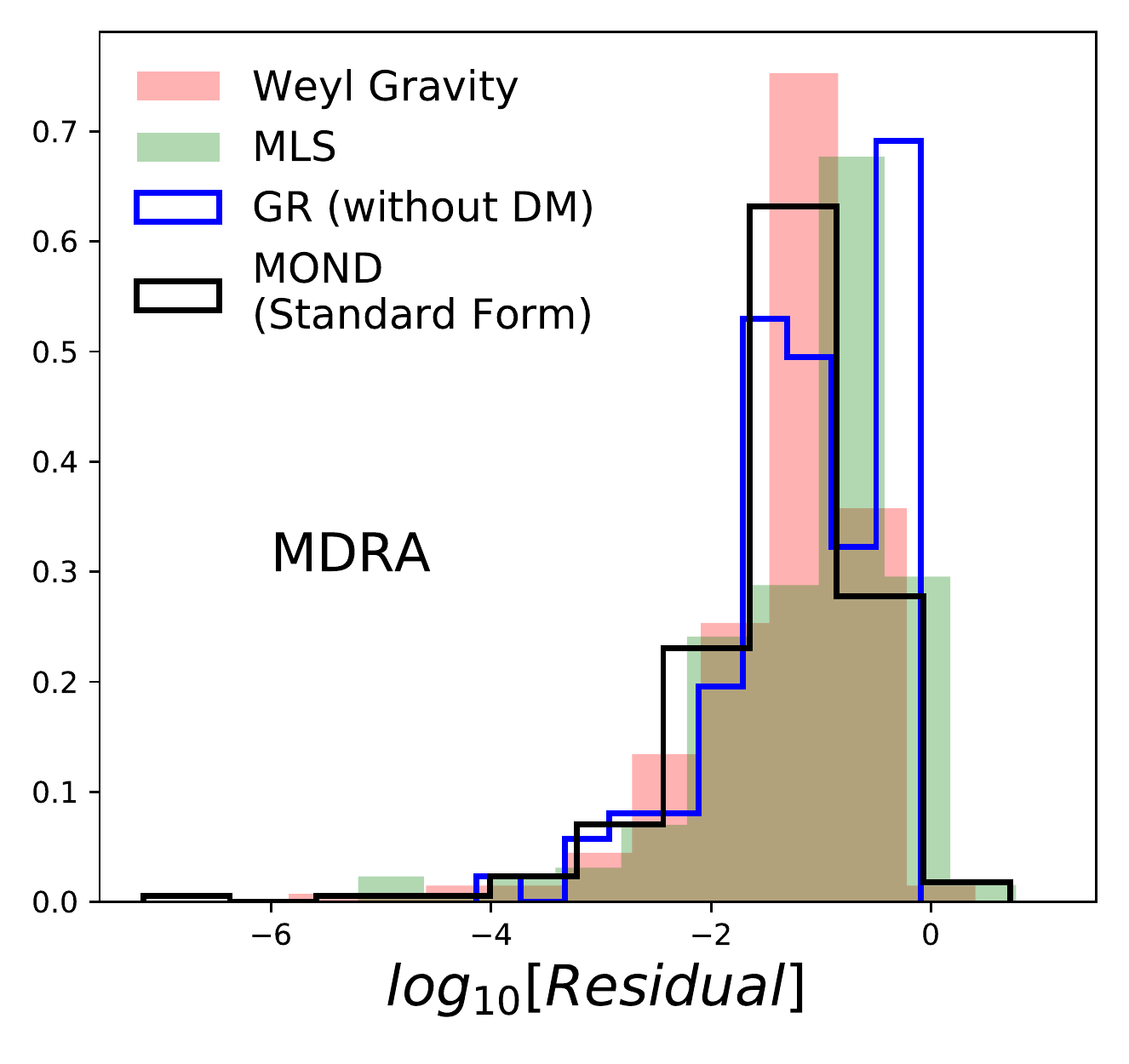} 
	\end{center} 
	\caption{\textbf{Histogram of the residuals between the inferred mass discrepancies and the predicted mass discrepancies in GR (without DM), Weyl gravity, MOND and RAR scaling. Color codes are given in the legend. Details are in the texts.}}
	\label{fig4}
\end{figure}
To understand the goodness-of-fit for different theories, we have plotted the residuals between data and model in Fig \ref{fig1} (lower panel) as a function of radial distances and expected Newtonian acceleration in loglog scales. For convenience, we use the following definition for residuals:
\begin{equation}
Residual = (Data - Model)^2/Data^2.
\label{residual}
\end{equation}
We choose this particular definition of residual for two reasons. First, the residuals are always positive and, thus, can easily be plotted in log-scale. This is necessary as the centripetal acceleration data span from $10^{-12}$ $m/s^{2}$ to $10^{-8}$ $m/s^{2}$. Second, taking only absolute difference between the observed data and predicted values can erroneously imply that a model having smaller differences in the high-acceleration regime is better than other models. To eliminate such possibility, we use a relative residual. Smaller values of residual indicates a better match between observed centripetal acceleration and the predicted accelerations in different theories of gravity. We find that GR (without DM) produces systematically larger error as distance increases (and acceleration decreases). However, at the high-acceleration regime, residuals for GR (without DM) is comparable to other theories in question. Finally, we plot the histograms of residuals for different theories in Fig \ref{fig2}. Residual histogram for GR (without DM) peaks at a larger value whereas MOND and Weyl gravity peaks overlap. The latter two histograms also exhibit longer tails in the lower end of residual values. RAR scaling produces residuals slightly larger than MOND and Weyl gravity.
\begin{table}
	\centering
	\caption[Milky Way mass model]{\textbf{Reduced chi-square values as goodness-of-fits for different theories of gravity and RAR scaling law. No dark matter is assumed. (Section \ref{sec4a} in text)}}
	\label{T2}
	\begin{tabular}{c c}
		\hline 
		\hline
		&$\chi^{2}/dof$ \\
		General Relativity (GR) without dark matter &7.56\\
		MOND (Standard Form) & 5.90\\
		Weyl Conformal Gravity & 6.11\\
		Radial Acceleration Relation / MLS 2016 & 5.71\\
		\hline 
		\hline
	\end{tabular} 
\end{table}
\begin{figure*}
	\begin{center}
		\includegraphics[scale=0.55]{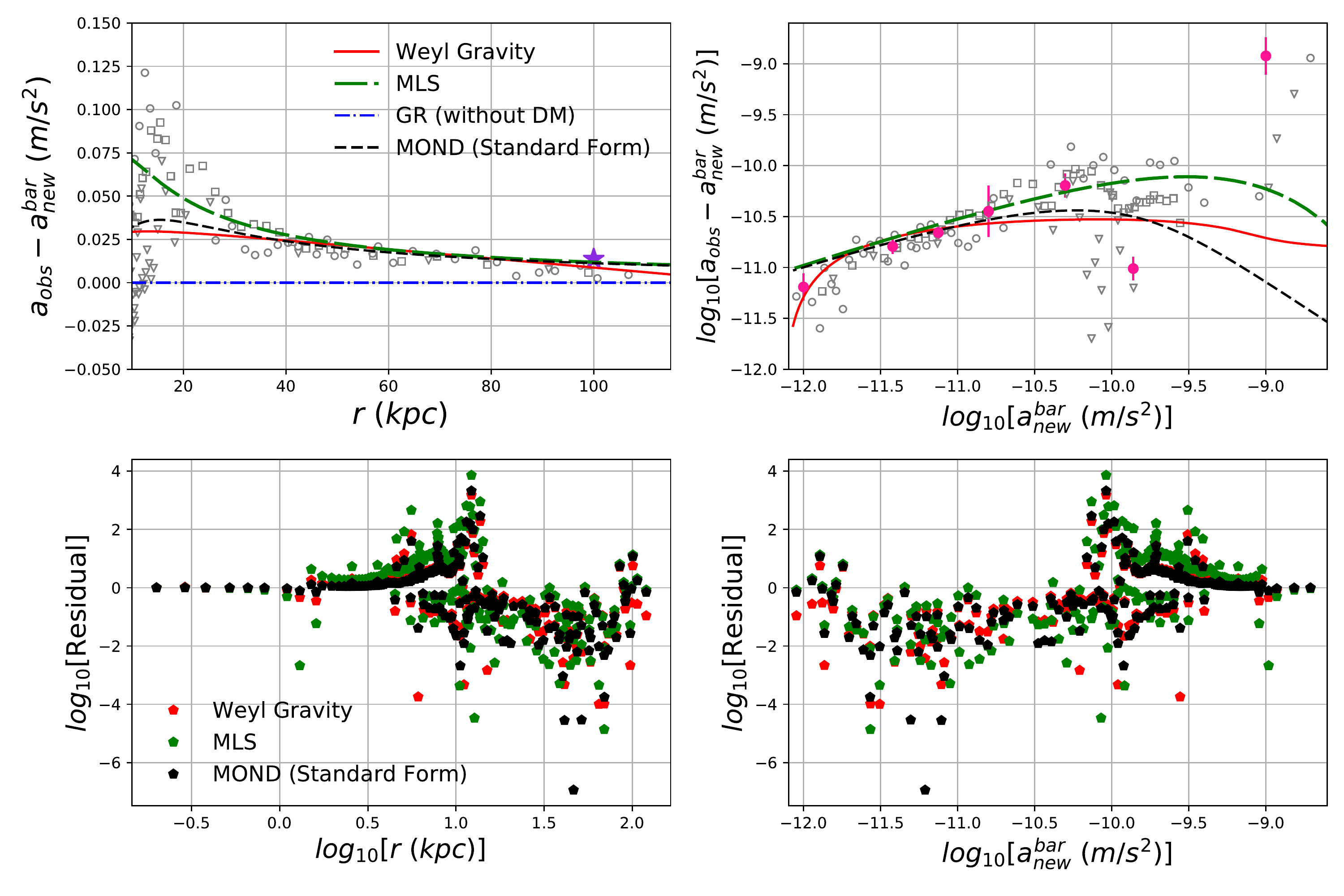} 
	\end{center} 
	\caption{\textbf{\textit{Upper Left: } Observed halo acceleration as  a  function  of  radial  distances from the galactic center. Violet star denotes the constant acceleration $\gamma_0 c^2/2$ in Weyl gravity. \textit{Upper Right: } loglog plot of observed halo acceleration as a function of Newtonian expectation due to baryons. Predicted profiles in GR (without dark matter), Weyl gravity, MOND and RAR scaling are then superimposed in both panels. Binned data plotted in pink circles. \textit{Lower Left: } Residuals [for GR (without dark matter), Weyl gravity, MOND and RAR scaling] as a function of radial  distances from the galactic center in log-log scale. \textit{Lower Right: } Residuals as a function of Newtonian expectation due to baryons. Color codes are given in the legend. Details are in the texts.}}
	\label{fig5}
\end{figure*}

\subsection{Mass Discrepancy-Radial Acceleration (MDRA) Relation and Modified gravity}
We now compute the (Newtonian) dynamical mass as a function of the the radial distances from the galactic center. The dynamical mass can directly be obtained as $M_{dyn}=a_{obs}r^{2}/G$. Similarly, one can write the baryonic mass in terms of the Newtonian acceleration due to baryons: $M_{bar}=a_{new}^{bar}r^{2}/G$. The ratio of the dynamical mass and baryonic mass is therefore same as the ratio of the observed acceleration and the expected Newtonian acceleration due to baryons: $M_{dyn}/M_{bar}=a_{obs}/a_{new}^{bar}$. This ratio is a measure of the `mass discrepancy' in a particular galaxy. In other words, it quantifies the amount of `missing mass' in a galaxy.  \\

In Figure \ref{fig3} (upper left), we plot the inferred ratio $M_{dyn}/M_{bar}$ ($=a_{obs}/a_{new}^{bar}$) as a function of the radial distances from the Milky Way center. We observe that the amount of missing mass (or the ratio of the observed and expected Newtonian acceleration due to baryons) increases as distance increases. The dashed blue indicates the scenario where observed acceleration equals to the expected Newtonian acceleration from baryons. We find that at larger distances MOND and RAR exhibits similar features whereas Weyl gravity profile departs from MOND/RAR profiles. These features become more prominent in Figure \ref{fig3} (upper left) where we plot the mass discrepancy as a function of the Newtonian acceleration due to baryons. We notice that, although MOND/RAR/Weyl gravity mass discrepancy profiles become similar to each other in the high acceleration regime (i.e. in interior of the galaxy), there is a difference between these predicted profiles and inferred mass-discrepancy data from YS12 \citep{sofue2}. We also plot the binned data for radial acceleration in pink circles. We find that Weyl gravity and MOND profile accounts for the binned data better than RAR scaling. In the lower panel of Figure \ref{fig3}, we plot the residual as a function of radial distances and expected Newtonian accelerations. It must be noted that the residual (defined in Eq. \ref{residual}) is same for RAR and MDRA. Histograms of residuals for different theories are shown in Fig \ref{fig4}.

\subsection{Halo Acceleration Relation (HAR) and Modified gravity}
The `halo acceleration' \citep{tian2019halo} is defined as the difference between the observed acceleration and the expected Newtonian acceleration due to baryons:
\begin{equation}
a_{h}=a_{obs}-a_{new}^{bar}.
\end{equation}
We now plot the radial variation of the `halo acceleration' in Figure \ref{fig5} (upper left). We find a scatter in data around zero in the interior of the galaxy (within $\sim$ 20 kpc from the galactic center) beyond which the data becomes almost independent of the radial distance. This feature is strikingly similar to the findings of  \cite{o2019radial} who observed that, beyond 10 kpc,  the difference between observed acceleration and expected Newtonian acceleration (due to baryons) in the cumulative sample of 207 galaxies is confined to  very narrow bracket which does not depend on radial distances anymore. Furthermore, the `halo acceleration' in this region systematically exhibits positive values hinting an underlying departure from Newtonian dynamics. We further find that Weyl gravity, MOND and RAR successfully capture this narrow band beyond 20 kpc. However, the inner region continues to be problematic for these theories/scaling to explain well. \\

It is important to point out that the asymptotic behavior of RAR, MOND and Weyl gravity profile have some subtle differences. In the low acceleration regime (i.e. for larger $r$), RAR goes as: $a_{MLS} \propto (a_{new}^{bar})^{(1/2)}$. Thus, the `halo acceleration' $a_{h,MLS}$ $\propto (a_{new}^{bar})^{(1/2)}-a_{new}^{bar}$. As, for larger $r$, $a_{new}^{bar}\rightarrow0$,  $a_{h,MLS}$ also goes to zero. Similarly, for the standard form of MOND, both $a_{MOND}$ and the difference between $a_{MOND}$ and $a_{new}^{bar}$ goes to zero in the lower acceleration limit. However, for the Weyl gravity, the asymptote takes the following form:
\begin{equation}
a_{weyl}=\frac{\gamma_0 c^{2}}{2} - \kappa c^{2} r.
\label{weylhar}
\end{equation}
Therefore, the acceleration becomes almost constant when the quadratic term is negligible. For larger distances from the galactic center, however, the negative quadratic term becomes significant such that $a_{weyl}$ (and consequently halo acceleration in Weyl gravity) approaches zero faster than MOND and RAR (Figure \ref{fig5}; upper left and upper right). Such subtle features can in principle be used in future tests of modified gravity theory with RAR (or HAR). Eq. (\ref{weylhar}) further suggests that $a_{weyl}$ (and halo acceleration in Weyl gravity) at larger distances from galactic center will have a maximum value of $\frac{\gamma_0 c^{2}}{2}$ (denoted by a `star' in Figure \ref{fig5} upper left). Both observed halo acceleration data and the predicted Weyl gravity profile are found to comply with this upper bound.\\
\begin{figure}
	\begin{center}
		\includegraphics[scale=0.6]{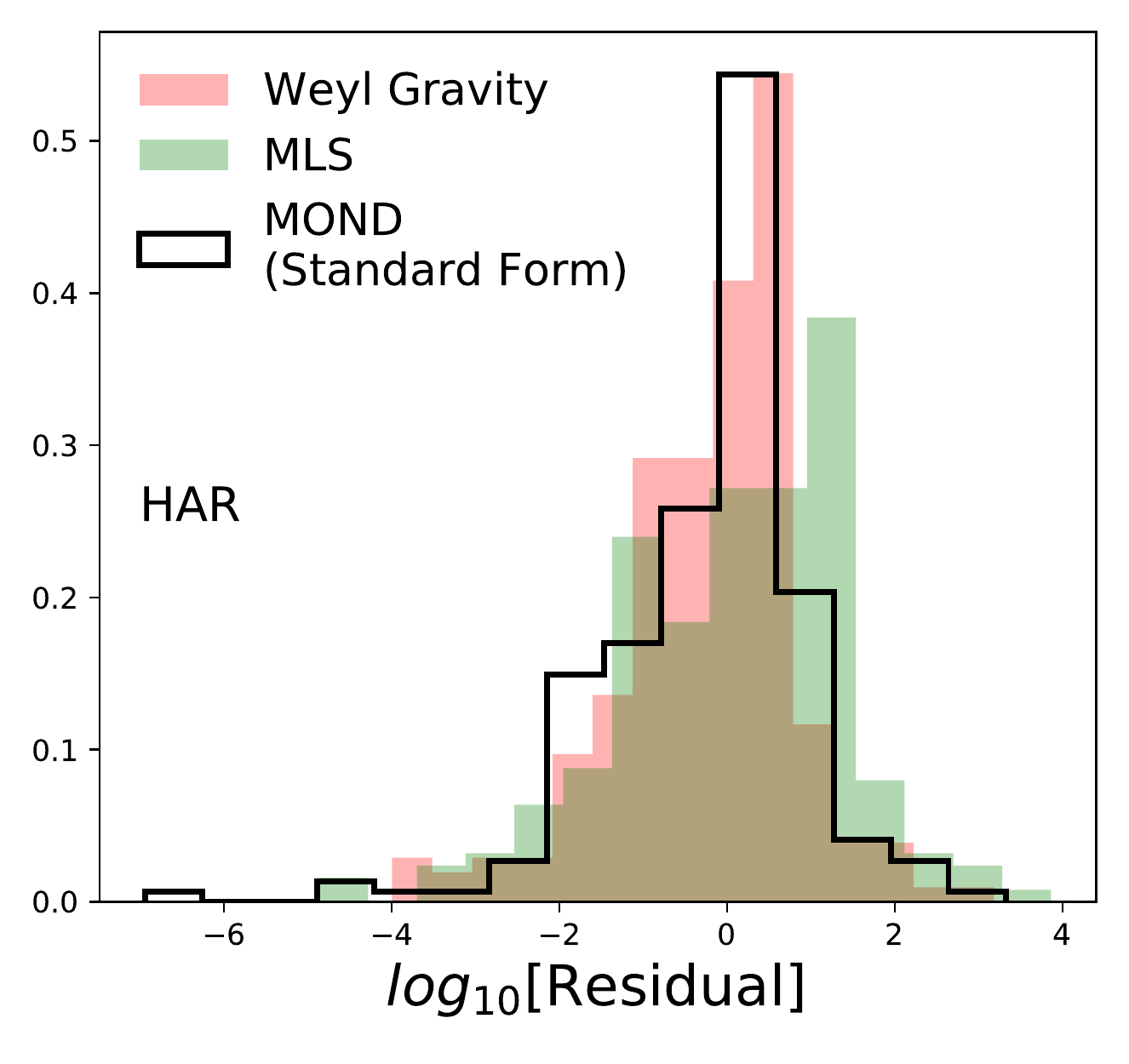} 
	\end{center} 
	\caption{\textbf{Histogram of the residuals between the observed halo accelerations and the predicted halo accelerations in GR (without DM), Weyl gravity, MOND and RAR scaling. Color codes are given in the legend. Details are in the texts.}}
	\label{fig6}
\end{figure}
To investigate this region more carefully, we now plot the halo acceleration data in log-log scale as a function of the Newtonian acceleration expected from baryons. We do not find any clear evidence for the existence of a maxima in $a_h$ as claimed by \cite{tian2019halo} (Figure \ref{fig5}; upper right). Binned halo acceleration data initially shows an uni-modal feature only then to increase in the interior of the galaxy. However, we find that casting the data into $a_h\textendash a_{new}^{bar}$ plane helps to discriminate between different theoretical models. For example, the expected profiles in Weyl gravity, MOND and RAR originating from the baryons in the Milky Way looks very similar to each other when plotted in the $a_{obs}^{bar}\textendash r$ plane or $a_{obs}\textendash a_{new}^{bar}$ plane or $a_h\textendash r$ plane. However, in the halo acceleration vs Newtonian acceleration (due to baryons) plane, they look strikingly different from each other. These differences could be exploited further to discriminate between different models. Interestingly, we find uni-modal feature in both MOND and RAR profiles while Weyl gravity curve does not show any such signature. Moreover, it is surprising to see that the high acceleration regime proves to be more vital when the question pops up: which model better explains the data? \\

Overall, we observe that Weyl gravity and MOND produces smaller residuals than RAR scaling (Figure \ref{fig6}). At this point, we note that the discrepancy between the data and expected profiles in Weyl gravity, MOND and RAR is considerably high in the high end of acceleration regime which, in general, corresponds to the innermost region of the galaxy (Figure \ref{fig5}; lower panels). One particular possibility is that the mass model, used to generate the expected modified gravity/RAR profiles, is not adequate in this region. That could be the case in the Milky Way as we ignore the effects of the presence of `holes' in the inner region of the gas disks \citep{mcmillan}. The effects of the black hole are also taken naively. These issues should be taken care of if one pursues a test of modified gravity theories with halo acceleration relation.  \\

We therefore conclude that RAR definitely gives a strong test for modified gravity theories and dark matter models. It would probably continue to be one of the zeroth order tests any modified gravity theory must pass at the galactic scale. However, HAR would enable us to formulate a precision test which will require finer knowledge about the mass model of a particular galaxy (the Milky Way for this work).
\section{\textbf{Concluding Remarks}}
\label{sec5} 
In this work, we have used the inferred acceleration data in the Milky Way obtained from different kinematic surveys \citep{sofue2,pijush,yh16} to test RAR and two popular modified gravity theories, MOND (standard form) and Weyl gravity. It must be noted that the RAR scaling proposed by McGaugh, Lelli  and Schombert (MLS) \citep{McGaugh} is in fact another form of MOND with different interpolating function. In that sense, this work tests Weyl gravity and two different versions of MOND. We have found that both the modified gravity theories in question as well as RAR can explain the radial acceleration data well. We further investigated whether representing the data in the form of halo acceleration (i.e. difference between observed and expected Newtonian acceleration due to baryons) yields anything extra. We have noticed that while the data in the $a_{obs}\textendash a_{new}^{bar}$ plane is unable to discriminate between different models or gravity and scaling laws, $a_{halo} \textendash a_{new}^{bar}$ plane gives a stronger test for them. We have further observed that, in the $a_{halo}\textendash a_{new}^{bar}$ plane, both the high acceleration and low acceleration regime becomes equally important for such tests. In our case, we demonstrated that, though in the low acceleration regime the predicted profiles in MOND, RAR and Weyl gravity reasonably agree with each other, their trajectory differs significantly in the high acceleration regime. We also note that the current uncertainties and inadequacy of mass models in the high acceleration regime (i.e. in the innermost part of the Milky Way) does not allow us to reach any strong conclusion. However, in future, as more accurate mass model becomes available, one can formulate precision tests for modified gravity theories (and dark matter models) against acceleration data in the $a_{halo} \textendash a_{new}^{bar}$ plane.\\

\noindent \textbf{Acknowledgement:}
We thank the referees for their thoughtful comments towards improving the paper. TI's research is supported by a Doctoral Fellowship at UMass Dartmouth and NSF grant PHY 1806665. TI thanks the Long-Term Visiting Fellowship Program at ICTS-TIFR during which the work began. KD would like to thank ICTP, Trieste for its Associateship Program. Computation has been carried out in the cluster CARNIE at Center for Scientific Computation and Visualization Research (CSCVR), University of Massachusetts (UMass) Dartmouth.

\end{document}